\magnification=\magstep1
\tolerance=1600
\font\fta=cmr10 scaled\magstep2
\font\ftb=cmsy10 scaled\magstep2
\font\ftc=cmmi10 scaled\magstep2
\centerline{\fta Observation of the Charge Symmetry Breaking}
\medskip
\centerline{\fta 
d+d{\ftb\char'041}\raise1ex\hbox{\rm 
4}He+{\ftc\char'031}\raise1ex\hbox{\rm 0}
Reaction Near Threshold}
\vskip.4in
\centerline{E.J. Stephenson,$^1$ A.D. Bacher,$^{1,2}$
C.E. Allgower,$^1$ 
A. G\aa rdestig,$^3$}
\centerline{C. Lavelle,$^1$ G.A. Miller,$^4$ H. Nann,$^{1,2}$ 
J. Olmsted,$^1$
P.V. Pancella,$^5$ M.A. Pickar,$^6$}
\centerline{J. Rapaport,$^7$ T. Rinckel,$^1$ A. Smith,$^8$ H.M.
Spinka,$^9$ and U. van Kolck$^{10,11}$}
\bigskip
\centerline{\it $^1$Indiana University Cyclotron Facility, Bloomington,
IN 47408}
\centerline{\it $^2$Department of Physics, Indiana University,
Bloomington, IN 47405}
\centerline{\it $^3$Indiana University Nuclear Theory Center,
Bloomington, IN 47408}
\centerline{\it $^4$Department of Physics, University of Washington,
Seattle, WA  98195}
\centerline{\it $^5$Physics Department, Western Michigan University,
Kalamazoo, MI 49008}
\centerline{\it $^6$Department of Physics and Astronomy, Minnesota
State University}
\centerline{\it at Mankato, Mankato, MN 56001}
\centerline{\it $^7$Department of Physics and Astronomy, Ohio
University, Athens, OH 45701}
\centerline{\it $^8$Physics Department, Hillsdale
College, Hillsdale, MI 49242}
\centerline{\it $^9$Argonne National Laboratory, Argonne, IL 60439}
\centerline{\it $^{10}$Department of Physics, University of Arizona,
Tucson, AZ 85721}
\centerline{\it$^{11}$RIKEN BNL Research Center, Brookhaven National 
Laboratory,
Upton, NY 11973}
\vskip.4in
\noindent ABSTRACT:
We report the first observation of the charge symmetry breaking
${\rm d}+{\rm d}\to{^4{\rm He}}+\pi^0$ reaction near threshold at
the Indiana University Cyclotron Facility.
Kinematic reconstruction permitted the separation of $^4{\rm He}+
\pi^0$ events from double radiative capture $^4{\rm He}+\gamma +\gamma$
events.  We measured total cross sections for neutral pion production
of $12.7\pm 2.2$~pb at 228.5~MeV and $15.1\pm 3.1$~pb at 231.8~MeV.
The uncertainty is dominated by statistical errors.
\vskip.4in
Charge symmetry is the approximate symmetry of the strong
interaction under a specific isospin rotation that interchanges
down and up quarks [1,2].
This symmetry is broken by the different masses of the down and
up quarks
($m_d>m_u$) and by their electromagnetic interactions.
The combination of these two mechanisms leads, 
for example, to
the neutron
being heavier than the proton.  Within the framework of chiral
effective field theory 
[3,4], additional experimental information on the
relative contributions of these two mechanisms to charge symmetry
breaking (CSB) must, in leading order,
come from pion-nucleon scattering.  Direct experimental evidence
is restricted to elastic scattering and charge exchange experiments with
low-energy charged pions where the interpretation is complicated
by corrections for the neutron-proton mass difference and
electromagnetic interactions between the pions and nucleons
[5,6].  
Reactions in which a $\pi^0$ is emitted after being created by one
nucleon and rescattered by a second are particularly clean.
One example of such a CSB process
is the measurement of a forward-backward
asymmetry in the cross section for the ${\rm n}+{\rm p}\to{\rm d}+
\pi^0$ reaction [7].

The ${\rm d}+{\rm d}\to{^4{\rm He}}+\pi^0$ reaction
violates charge symmetry because the $\pi^0$, whose
wavefunction is odd under the interchange of down and up quarks,
should not be produced from charge symmetry even (in
this case self-conjugate)
nuclear systems.  
The amplitude for CSB is weaker than a similar charge symmetry
conserving amplitude by about 1/300, roughly the ratio of the quark
mass difference to the nucleon mass.  This suggests a
${\rm d}+{\rm d}\to{^4{\rm He}}+\pi^0$ total cross section,
which depends only on a CSB amplitude, that is as small as
tens of picobarns.

Several searches for the ${\rm d}+{\rm d}\to{^4{\rm He}}+\pi^0$ reaction
have produced only upper limits 
[8].  A positive report at a deuteron
energy of 1.1~GeV 
[9] has been questioned because the experiment did
not clearly distinguish the photons from $\pi^0$ decay from photons that
could have been produced by the double radiative capture ${\rm d}+{\rm d}
\to{^4{\rm He}}+\gamma +\gamma$ process. 
This process is isospin allowed and predicted to appear at the
reported cross section [10].
A further search that can clearly distinguish between these two
reactions is therefore warranted.

We chose to look for the ${\rm d}+{\rm d}\to{^4{\rm He}}+\pi^0$
reaction just above its threshold at 225.5~MeV to avoid other pion
producing channels and to take advantage of the clean experimental
conditions afforded by the Indiana University Cyclotron Facility's
electron-cooled storage ring.  A $6^\circ$ bend located in one section
of the ring provided a site where $^4$He nuclei, produced in a narrow
forward cone just above threshold, could be separated from the 
circulating deuteron beam.  By placing a gas jet target sufficiently
upstream of the $6^\circ$ magnet, it became possible to cover a large
solid angle with two arrays of Pb-glass detectors that would be
selectively sensitive to photons from the target region.

\topinsert
\vbox to 0.1truein{\vss
\hbox to 6.5truein{
\includegraphics{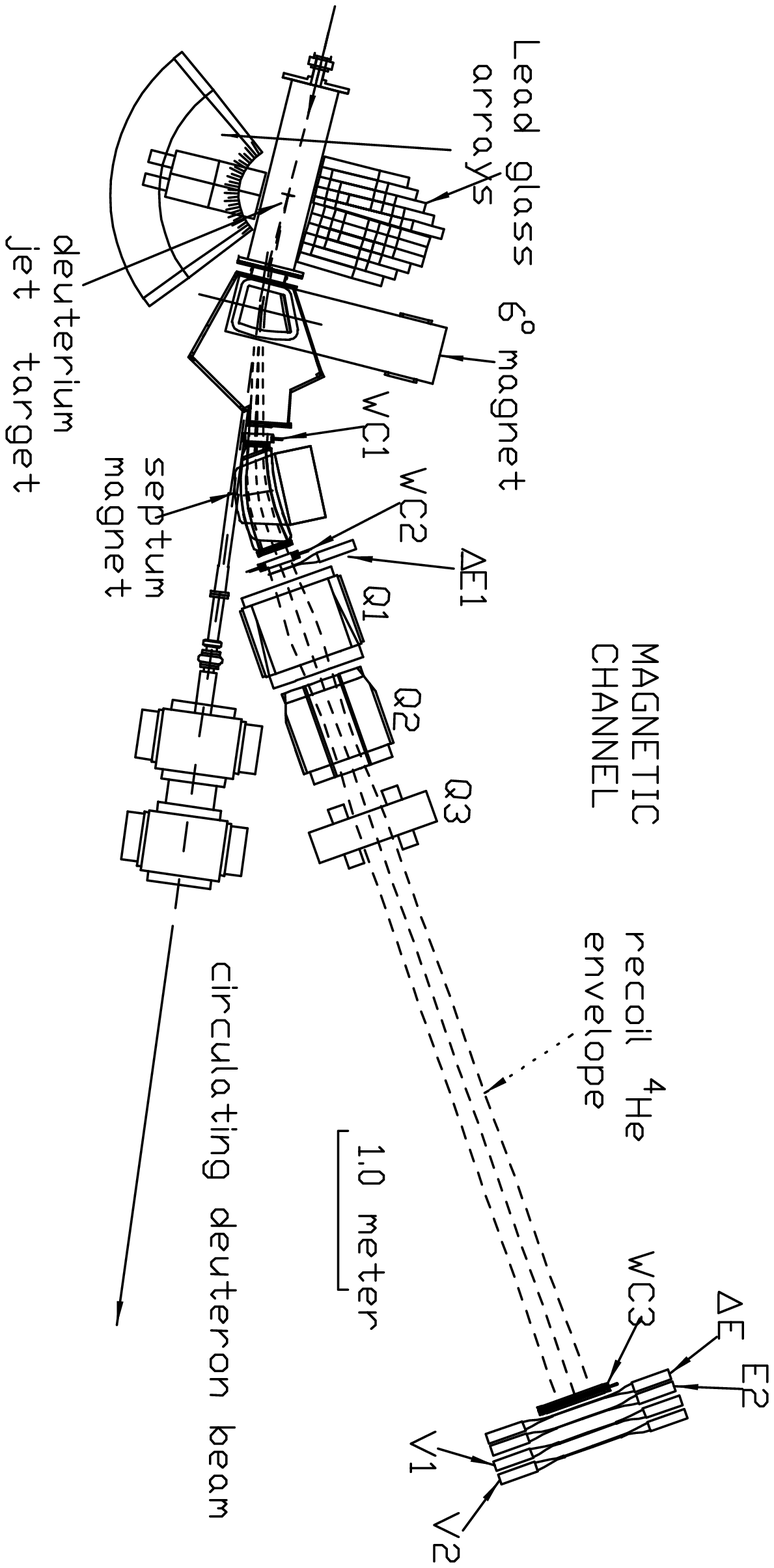}\hss}}
\vskip3.4truein
\noindent{\it Figure 1:~~Layout of the experimental setup showing the
target, approximate locations of the Pb-glass arrays, and the magnetic
channel in relation to a segment of the electron-cooled storage ring.
Quadrupole magnets (Q1, Q2, and Q3), wire chambers (WC1, WC2, and 
WC3), and scintillation trigger ($\Delta$E1, $\Delta$E2, and E) and
veto (V1 and V2) detectors are shown.  The luminosity detectors are
small and consequently are omitted here.}
\endinsert

The layout of the experiment is shown in Fig.~1, with 
the major features
being two arrays of Pb-glass detectors stacked on the left and right
sides of the gas target box, the $6^\circ$ separation magnet, and a
magnetic channel consisting of a septum magnet to steer the $^4$He
nuclei away from the downstream ring quadrupole magnets and a 
quadrupole triplet to confine the $^4$He nuclei within the acceptance
of the detectors at the end of a long, evacuated drift length.

For some $^4{\rm He}+\pi^0$ events, the $^4$He nucleus and both $\pi^0$
decay photons could be recorded.  Parts of the solid angle above and
below the target were required for differential pumping of the jet target
volume.
Kinematic constraints just above threshold require very similar
photon opening angles and energies for $^4{\rm He}+\pi^0$ and 
$^4{\rm He}+\gamma+\gamma$ events.
Separation of these two reactions
relied on measuring the $^4$He momentum vector
using time of flight in the
channel for the longitudinal component and scattering angle for the
transverse component.  The flight time was measured from the first
plastic scintillation detector ($\Delta$E1) 
to the second
($\Delta$E2) at the end of the channel
5.73~m away.  A third scintillator (E) in
which the $^4$He nuclei stopped was required for an event trigger along
with the absence of any signal in two additional scintillators (V1 and
V2).  Pulse-height correlations among the trigger scintillators
cleanly separated $^4$He events.
Pb-glass information was not used in the trigger.  The
scattering angle of the $^4$He was recorded by a multiwire proportional
chamber (WC1).  Two additional
chambers (WC2 and WC3) tracked each $^4$He nucleus through the channel.
From the $^4$He momentum and the assumption of a two-body final state,
it is possible to calculate the missing mass.  Double
radiative capture events produce a broad distribution of mass values
up to a kinematic limit that depends on the beam energy
rather than a peak at the $\pi^0$ mass.


The isospin-allowed ${\rm p}+{\rm d}\to{^3{\rm He}}+\pi^0$ reaction was
used to commission the
detector system.
The calculation of $^3$He (and later $^4$He) momentum from
channel time of flight used a model to describe the energy loss 
(scaled from Janni [11]) of the
$^3$He passing through the channel detectors and vacuum windows.
The time 
offsets for each photomultiplier associated with the
$\Delta$E1 and $\Delta$E2 detectors were adjusted empirically during
the subsequent data analysis.  For $^3$He energies
close to threshold and runs lasting several hours, a $\pi^0$ mass
resolution as small as ${\rm FWHM}=240$~keV was obtained.

For ${\rm p}+{\rm d}\to{^3{\rm He}}+\pi^0$ events in which both $\pi^0$
photons were recorded, photon energies were taken to be the sum of
Pb-glass energies starting with the detector recording the highest
energy deposition and including its eight nearest neighbors.
With summed energies above a threshold
chosen to remove most random events and 
with Pb-glass timing in the correct
range relative to the channel trigger, the 
measured efficiency for detecting the
two $\pi^0$ decay photons was $0.360\pm 0.001$.
Simulations using the GEANT
Monte-Carlo program library [12] and the known $\pi^0$
angular distribution [13]
reproduced the Pb-glass spectral energy
shape and agreed with the measured efficiency to $\pm 0.01$
This simulation was used to obtain the {\it changes} in Pb-glass
efficiency between the ${\rm p}+{\rm d}\to{^3{\rm He}}+\pi^0$
commissioning run and production running for ${\rm d}+{\rm d}\to
{^4{\rm He}}+\pi^0$ due to alterations of the kinematics and
center-of-mass angular distribution (assuming the ${\rm d}+{\rm d}\to
{^4{\rm He}}+\pi^0$ cross section to be isotropic).  Gains for the
Pb-glass detectors were calibrated using cosmic ray muons.  This
calibration operated continuously during commissioning and production
running.

The size of the $^3$He opening angle cone for ${\rm p}+{\rm d}\to
{^3{\rm He}}+\pi^0$ calibrated the incident proton energy.
Several determinations at slightly different energies were combined by
using the frequency of the RF voltage that maintained the beam bunching
to calculate the circumference of the storage ring, yielding a value of
$86.786\pm 0.003$~m that was consistent over time.  This calibration
was used subsequently for the deuteron beam energy.

The deuterium target was made by directing deuterium gas from a glass
nozzle cooled to about 40~K across the circulating storage ring
beam ($\sim 2$~mm wide).  
The gas flow was adjusted so that the beam lifetime was close to
100~s with the RF voltage and electron cooling on, a value that gave the
largest data production rate.
The circulating beam 
current was as large
as 2~mA.

The luminosity (product of beam flux and intercepted target thickness)
was determined through a separate calibrated monitoring system.
For this we used ${\rm d}+{\rm d}$ elastic scattering in the vicinity
of $\theta_{c.m.}=90^\circ$ by placing two scintillator detectors at
$44^\circ$ on either side of the beam.  The product of ${\rm d}+{\rm d}$
differential cross section and solid angle for this system was measured
in a separate experiment using a molecular HD target.  Additional
scintillators were added to observe ${\rm d}+{\rm p}$ elastic
scattering.  Using known ${\rm d}+{\rm p}$ cross sections and the 
equality of ${\rm d}+{\rm p}$ and ${\rm d}+{\rm d}$ luminosities for an
HD target, we obtained the ${\rm d}+{\rm d}$ calibration.  Cross sections
for ${\rm d}+{\rm p}$ elastic scattering in our energy range have
recently become available 
[14] with an absolute normalization error of
5.5\%.  The scintillation detectors provided particle
identification information to separate breakup
and other backgrounds.  An additional position-sensitive
silicon detector used to record recoil nuclei from small-angle
scattering provided a measurement of the jet target profile as it
intercepted the beam.  This information was used to make corrections to
the scintillator acceptance geometry for both the ${\rm d}+{\rm p}$
and ${\rm d}+{\rm d}$ elastic scattering processes, and to track changes
between calibration and production running.  The 
average luminosity during
production was $2.9\times 10^{31}$~/cm$^2$/s.

Candidate ${\rm d}+{\rm d}\to{^4{\rm He}}+\pi^0$ events were required
to have the correct pulse height in the three trigger scintillators
($\Delta$E1, $\Delta$E2, and E), usable wire chamber information, and
photon signals in {\it both} the left and right Pb-glass arrays with
an energy sum above threshold and timing
coincident with $^4$He events in the channel.  For
each candidate event the missing mass was calculated.  The
results were
accumulated into a spectrum, as shown in Fig.~2.  Investigations
of the quality of the data set showed that random background was
essentially removed ($\leq 1$~event/spectrum); thus we should interpret
these results as arising only from $\pi^0$ production and double 
radiative capture.

\midinsert
\vbox to 0.1in{\vss
\hbox to 6.5truein{
\includegraphics{csbfig2.ps}\hss}}
\vskip0.7in
\hsize=2.5truein
\noindent{\it Figure 2:~~Historgrams of the candidate
events at the two deuteron
bombarding energies as a function of their missing mass value.  The
smooth curves show the reproduction of these histograms with a Gaussian
peak and a continuum.}
\vskip1.2in
\endinsert

Production running started at 228.5~MeV, an energy chosen because the
half angle of the
$^4$He forward cone 
($1.2^\circ$) was expected to fit safely inside the
channel acceptance.  Over the much longer ${\rm d}+{\rm d}\to{^4{\rm He}}
+\pi^0$ production running times, drifts in the time-of-flight 
measurement became more of a problem, despite efforts to track timing
changes by monitoring other particles going along the channel.
The width of the missing mass peak
was ${\rm FWHM}=510$~keV, as seen in the
top panel of Fig.~2.  This meant that the peak was not well separated
from the $^4{\rm He}+\gamma +\gamma$
kinematic upper limit of 136.4~MeV.  Below the peak, the flux
we would attribute to double radiative capture is attenuated by the
acceptance boundaries of the channel, increasing the ambiguity of its
shape.  The decision was made to complete production running at the
higher energy of 231.8~MeV ($1.75^\circ$ cone) with a kinematic
endpoint of 138.0~MeV
even though an additional 10\%
of the $\pi^0$ events would be lost in the channel.  
The result shown in the
lower panel of Fig.~2 still has a broad peak 
(${\rm FWHM}=660$~keV) but there is now
a clear
distribution of double radiative capture events on either side.

The 
spectra of Fig.~2 were modelled with a Gaussian peak and a 
continuum.  The continuum shape was obtained from the distribution 
in missing mass of
all $^4$He events (rate about $10^3$ higher
than for candidate events alone) reduced by the ratio of the
calculated double radiative capture cross section to the phase space
value.  These $^4$He events without coincident
photons, which may have originated from
(d,$^4$He) reactions on residual gas and storage ring structures, were
broadly distributed in energy and angle, and were used as an estimate of
the product of phase space and channel acceptance.  
The continuum curves in Fig.~2
are a smooth representation of this shape.  The centroids of the Gaussian
peaks have a fitting error of less than 60~keV and are consistent with
the $\pi^0$ mass.  Events in the peak appear to be isotropically
distributed in the center of mass as shown by their distribution
in scattering angle and time of flight.

The 66 and 50 $^4{\rm He}+\pi^0$
events recorded at the two energies of 228.5 and 231.8~MeV
lead to total cross section values of $12.7\pm 2.2$~pb and $15.1\pm
3.1$~pb respectively, including a 6.6\% normalization
error for all systematic
effects.  Corrections at the two energies
included Pb-glass efficiency (0.34 and 0.32), 
trigger
losses (0.94 and 0.96 for random vetoes), 
system livetime (0.95 and 0.94),
wire chamber efficiency (0.93 and 0.95), 
and other channel losses from
acceptance and multiple scattering (0.95 and 0.81).  These cross 
sections
are consistent with being proportional to $\eta =p_\pi /m_\pi$ with a
combined slope of $\sigma_{\rm TOT}/\eta =80\pm 11$~pb.  The integral
of the double radiative capture process for the 2~MeV range just below
the kinematic limit is $6.9\pm 0.9$~pb and 
$9.5\pm 1.4$~pb, values that
are more than double predictions
[15].

These results provide the first unambiguous measurement of the 
${\rm d}+{\rm d}\to{^4{\rm He}}+\pi^0$ cross section as well as that
for the competing double radiative capture process.  A detailed
interpretation of our results will require a careful consideration
of different CSB reaction mechanisms.  On
the basis of chiral power counting arguments, the leading terms are
expected to be [16] those related to $\pi -\eta$ mixing
[17] and CSB $\pi +{\rm N}$ scattering
[3].  Preliminary plane wave calculations [16]
suggest that $\pi -\eta$ mixing may be important because
all four nucleons contribute coherently, especially if the $\eta$
is produced by two nucleons via the exchange of a heavy meson, as
in ${\rm p}+{\rm p}\to{\rm p}+{\rm p}+\pi^0$ [18].  The
effects of pion scattering are also expected to be enhanced by
the initial-state ${\rm d}+{\rm d}$ interactions [16].
Preliminary estimates show that electromagnetic effects are very
small.

Both $\pi -\eta$ mixing and CSB $\pi +{\rm N}$
scattering are also important in producing an asymmetry about
$90^\circ$ in the ${\rm n}+{\rm p}\to{\rm d}+\pi^0$ reaction
[4], with the latter determining the sign of the asymmetry.
Calculations now underway should allow the extraction of both
processes from the combined experimental results.  Since both depend
on the quark mass difference, experimental information that
separates each contribution will greatly enhance efforts to relate
the hadronic matrix elements to the underlying QCD theory.
\bigskip
We would like to thank V. Anferov, G.P.A. Berg, B. Chujko,
J. Doskow, G. East, C.C. Foster, W. Fox, D. Friesel, 
T. Hall, C.J. Horowitz, A. Kuznetsov,
V. Medvedev, H.O. Meyer, D. Patalahka, R.E.
Pollock, A. Prudkoglyad, S. Shastry, T. Sloan, K. Solberg,
J. Vanderwerp, and B. von Przewoski for
their contributions to the preparation of this experiment.  Our 
work was supported in part by the National Science Foundation grants
PHY-97-22538, PHY-98-02872,
and PHY-00-70368, the Department of Energy grant
DE-AC02-98CH10886, the Department of Energy OJI award
DE-FG03-01ER41196, RIKEN, Brookhaven National Laboratory, and an
Alfred P. Sloan Fellowship.
\vskip.3in
\item{1.} G.A. Miller, B.M.K. Nefkens, and I. \v Slaus,
Phys.\ Rep.\ {\bf 194}, 1 (1990).
\item{2.} G.A. Miller and W.T.H. van Oers, in {\it 
Symmetries
and Fundamental Interactions in Nuclei}, eds.\ W.C. Haxton and E.M.
Henley (World Scientific, Singapore, 1995) p.\ 127; and references
therein.
\item{3.} U. van Kolck, Few Body Systems, Suppl.\ {\bf 9},
444 (1995).
\item{4.} U. van Kolck, J.A. Niskanen, and G.A. Miller,
Phys.\ Lett.\ B {\bf 493}, 65 (2000).
\item{5.} W.R. Gibbs, Li Ai, and W.B. Kaufmann, Phys.\ Rev.\ 
Lett.\ {\bf 74}, 3740 (1995).
\item{6.} C. Matsinos, Phys.\ Rev.\ C {\bf 56}, 3014 (1997).
\item{7.} 
A.K. Opper and E. Korkmaz, TRIUMF experiment 704 proposal.
\item{8.} 
J. Banaigs {\it et al.},\ Phys.\ Rev.\ Lett.\ {\bf 58}, 1922
(1987); and references therein.
\item{9.} 
L. Goldzahl, J. Banaigs, J. Berger, F.L. Fabbri, J. H\"ufner,
and L. Satta,
Nucl.\ Phys.\ {\bf A533}, 675 (1991).
\item{10.} 
D. Dobrokhotov, G. F\"aldt, A. G\aa rdestig, and
C. Wilkin,
Phys.\ Rev.\ Lett.\ {\bf 83},
5246 (1999).
\item{11.} 
J.F. Janni, At.\ Data and Nucl.\ Data Tables {\bf 27},
150 -- 529 (1982).
\item{12.} GEANT program library, 
http://wwwinfo.cern.ch/asdoc/geant.html3/geantall.html.
\item{13.} M.A. Pickar {\it et al.},\ Phys.\ Rev.\ C {\bf 46},
397 (1992).
\item{14.} K. Ermisch, Ph.D. thesis, Rijkuniversiteit
Groningen, 2003.
\item{15.} A. G\aa rdestig, private communication.
\item{16.} A. Fonseca, A. G\aa rdestig, C. Hanhart, C.J.
Horowitz, G.A. Miller, A. Nogga, and U. van Kolck,
private communication.
\item{17.} S.A. Coon and B.M. Preedom, Phys.\ Rev.\ C {\bf 33},
605 (1986).
\item{18.} T.-S.H. Lee and D.O. Riska, Phys.\ Rev.\ Lett.\
{\bf 70}, 2237 (1993).
\vfill
\end